# ENERGY AND LINK QUALITY BASED ROUTING FOR DATA GATHERING TREE IN WIRELESS SENSOR NETWORKS UNDER TINYOS - 2.X

A. Sivagami[1], K. Pavai[2], D. Sridharan[3] and S.A.V. Satya Murty[4]

[1,2,3]Department of Electronics and Communication Engineering, Anna University, Chennai, India
[1]siva_psg69@yahoo.co.in, [2]pavai_me@yahoo.com, [3]sridhar@annauniv.edu
[4]Head, Computer Division, Indira Gandhi Centre for Atomic Research, Kalpakkam, India
satya@igcar.gov.in

## ABSTRACT

*Energy is one of the most important and scarce resources in Wireless Sensor Networks (WSN). WSN nodes work with the embedded operating system called TinyOS, which addresses the constrains of the WSN nodes such as limited processing power, memory, energy, etc and it uses the collection Tree Protocol (CTP) to collect the data from the sensor nodes. It uses either the four-bit link estimation or Link Estimation Exchange Protocol (LEEP) to predict the bi directional quality of the wireless link between the nodes and the next hop candidate is based on the estimated link quality. The residual energy of the node is an important key factor, which plays a vital role in the lifetime of the network and hence this has to taken as one of the metric in the parent selection. In this work, we consider the remaining energy of the node as one of the metric to decide the parent in addition to the link quality metrics. The proposed protocol was compared with CTP protocol in terms of number of packets forwarded by each node and packet reception ratio (PRR) of the network. This work was simulated in TOSSIM simulator and the same was tested in Crossbow IRIS radio test bed. The results show that our algorithm performs better than CTP in terms of load distribution and hence the increased lifetime*

## 1. INTRODUCTION

Wireless Sensor Networks (WSN) consist of potentially large tiny sensor nodes, which are capable of sensing the parameters of interest, processing the data locally and communicating the processed information over the radio. WSN networks are having huge potential applications in the area of environmental monitoring, military surveillance and reconnaissance, structural monitoring, habitat monitoring, health monitoring and home automation. The main aim of the WSN nodes is to sense, process the parameter of interest at its close proximity to the place where it got generated, and transmit in wireless mode to the collection centre at low cost. Also, it should reach any type of unreachable terrains like huge hills and deep valleys, dense forests and narrow gaps in machineries, etc. Due to these objectives, the nodes are designed to be tiny and also of low cost. Hence, the sensor nodes are designed to have a low processing power, low memory and use low power radio. The nodes are mostly unattended hence the battery power is more precious [1][2]. The conservation of the battery power is the most challenging task in the protocol design and this can be done at any layer.

TinyOS is a free and open source component-based event driven operating system, which addresses these issues of the WSN motes. TinyOS is implemented using nesC (Network Embedded System C) language, which supports the event based concurrent model of TinyOS [3]. The applications are developed from tiny reusable components, which are specific to the application, and the size of the code is in the order of kilobytes. TinyOS provides components for packet communication, sensing, scheduling, routing and medium accessing etc. The routing protocols supported by this operating system is Collection Tree Protocol (CTP), it estimates the quality of the link based on either one of the link estimator protocols such as LEEP or four-bit link estimator.

The quality of the link is measured with two parameters namely Received Signal Strength Indicator (RSSI) and Link Quality Indicator (LQI). Many of the radio chips like CC2420, RF230, etc provide the reading for both parameters and any one of the parameters can be taken as a metric to estimate the link quality. RSSI is the measure of the signal power from the received packet over 8 symbols whereas the LQI is the measure of the chip error rate over 8 bit period after the start of frame delimiter. The LQI has good correlation with Packet Reception Ratio (PRR) and thus it gives the better estimate of link quality over time [4]. Also, the LQI is not a fixed quantity, even though the distance between the nodes is unchanged, it varies over the time due to fading. The causes for fading are the multipath reflections from the obstacles present in the surrounding and the interference from other sources. Hence, the LQI is the important metric to decide the next hop and it should be calculated dynamically. The LEEP protocol measure the link quality based on the packet reception ratio (PRR), but not considering the packet acknowledgement, results in increased retransmissions. The four-bit link estimator considers the information from physical, link and network layer information to find the link quality [5]. Hence the Collection Tree Protocol (CTP) based on the four-bit link estimator is considered in our work. The CTP protocol decides the parent node merely based on the link quality and thus the nodes with good quality link will always be selected as the parent candidate. It is apparent that a node with good wireless link will involve in more communication and be drained out quickly. These nodes will be exhausted soon and the network will be disconnected. Thus the balancing the traffic among the nodes is necessary and this can be achieved by considering the residual energy of the node as one of the metric in the routing strategy. This approach will choose the parent by considering the varying nature of the wireless link as well as the residual energy of the node and thus improves the lifetime of the network.

This algorithm is simulated with TOSSIM, which is a discrete event simulator for TinyOS applications [6]. The same algorithm is tested in the test bed consisting of nine IRIS motes, which are WSN motes from Crossbow Technology. It is based on Atmega 1281 microcontroller and having 8KB RAM, 128KB programmable memory and 512KB flash memory. It uses ATRF230 radio chip for communication [7]. This mote is supported under TinyOS2.x, which is the de facto standard development platform for the resource constrained embedded Sensor Network.

The rest of the paper is organized as follows: Section 2 gives the survey of existing routing protocols, Section 3 describes the proposed protocol section 4 elaborates the implementation of the proposed algorithm in the simulator and the testbed section 5 gives the performance evaluation of the proposed protocol. Finally Section 6 gives the conclusion.

## 2. RELATED WORK

Many energy based routing protocols are proposed in the literature for wireless sensor networks but very few proposals are addressing the real time deployment of the wireless sensor networks. There are varieties of flavours of WSN nodes available in market and each one has its own challenges. The following paragraphs give the review of some routing protocols, which are based on the link quality and the energy for the real world WSN nodes.

### 2.1. RLQ: Resource Aware and Link Quality Based Routing Metric

The RLQ routing protocol for wireless sensor and actor networks (WSAN) proposed in [8] is based on the quality of the link, energy and also the heterogeneity of the actor networks. The routing decision is based on the link cost which is the sum of normalized energy cost for the transmitter and receiver. The normalized energy cost depends on the energy consumption for transmission and reception, residual energy and the link quality. The weighing factors x and y are used to change the routing decision. If both are zero, the routing decision is based on the minimum hop count. If x=1 and y=0, the minimum total energy consumption path is the shortest path. If both x and y are 1, routing metric is based on both link quality and the residual energy of the node. This algorithm was tested in the test bed consisting of 21 Tmote sky nodes and the performance of RLQ is tested with shortest path algorithm, and LQI in terms of PRR, throughput and lifetime of the network.

### 2.2. SHRP (Simple Hierarchical Routing Protocol)

The SHRP protocol [9] aims at the reliable data delivery in energy efficient manner. It is a proactive protocol in which the routes are selected based on the battery capacity and the link quality. The SHRP architecture uses TinyOS protocol SP, in which the parent node selection is based on the information from the link layer. The SHRP protocol eliminates the nodes, which are having either LQI, or RSSI value of the link is below a minimum threshold or the nodes that do not have enough battery to execute Minimum Task Cycle (MTC), from the neighbour table. The protocol chooses the route that has maximum energy among all possible routes to the sink with minimum hop distance. The algorithm was tested with Tmote Sky motes for different topologies like star, mesh, tree etc and the performance parameter considered in this protocol is convergence time. The protocol is compared with Hop to Sink Protocol (HTS) and the authors claimed the proposed protocol has better convergence time.  Since, the total energy cost for the whole path is taken as the metric, which will not address the issue that if, any node in the path having less energy metric.

### 2.3. LQER(Link Quality Estimation based Routing Protocol)

The LQER protocol [10] is a routing protocol with the objective of reliable data delivery in an energy efficient manner. LQER protocol makes the path selection based on the historical states of link quality after minimum hop field is established. In this paper, a dynamic window concept (m; k) is used to record the link historical information and which integrates the approach of minimum hop field. Here ‘m’ is the number of successful transmission over the link for the window size of ‘k’ previous transmissions.

It uses Minimum Hop field establishment algorithm to find the minimum hop neighbours using flooding approach and each node will have the list of forwarding neighbours. Each node chooses its next hop candidate from this list, which are having best m/k value.

The LQER protocol is simulated in WSNSim for Mica2 platform and the performance parameters considered are energy efficiency and the packet success rate. The authors claim that the proposed algorithm performs better in terms of energy efficiency, packet loss rate and scalability.

### 2.4. Efficient and Multi-path protocol for Wireless Sensor Network

This protocol [11] is designed for ensuring the QoS requirement of multimedia transmission such as throughput, delay etc. It is based on directed diffusion, which is scalable and uses single energy efficient path for data transmission. In multipath routing, disjoint multiple paths are chosen which are based on the link quality and the delay incurred in the path.

The path cost for the selection of forwarding candidate is based on cumulative path_ETX and path_Delay. To get high throughput and low delay paths, the path cost is defined as

$$Path_Cost = Path_ETX^{\alpha} * Path_Dealy^{\beta}$$

The ETX based link quality is estimated from SNR value of the received packets and is calculated according to [12] and is given by $ETX = 1/(d_f * d_r)$ where $d_f$ and $d_r$ are the forward and reverse packet delivery ratio of the particular link.

It is simulated in NS2 and provides high throughput and less delay, which are the requirement for the high quality video transmission

## 3. Energy And Link Quality Based Routing Tree (ELQR)

The quality of the wireless link between any two nodes is not static, it varies over time that depends on channel fading. The channel fading is due to multipath transmission or shadowing and it is time dependent. The time varying nature of the channel quality might have the impact on the bit error rate (BER) of the received packet and a node receives a packet with more bit error will be discarded, leads to packet retransmission. In energy-constrained network like WSN, the packet retransmission plays an important role in the lifetime of the network. To enhance the lifetime of the network, the packet retransmissions should be minimized by sending packets through the link with good channel characteristics. The lifetime of the network is further increased by distributing the traffic load among the nodes, which are having good quality wireless links. Hence, the lifetime maximization is achieved by reducing the number of retransmissions and load balancing. The data-gathering tree is constructed based on the quality of the link and the residual battery energy of the node, which is dynamic over time. In this proposed protocol ELQR, the data-gathering tree is constructed based on these two metrics. This protocol is designed for the real world deployment of the wireless sensor networks consists of IRIS motes. Our proposed algorithm is based on the CTP protocol with four-bit link estimator. The architecture of the proposed algorithm is shown in the figure1.

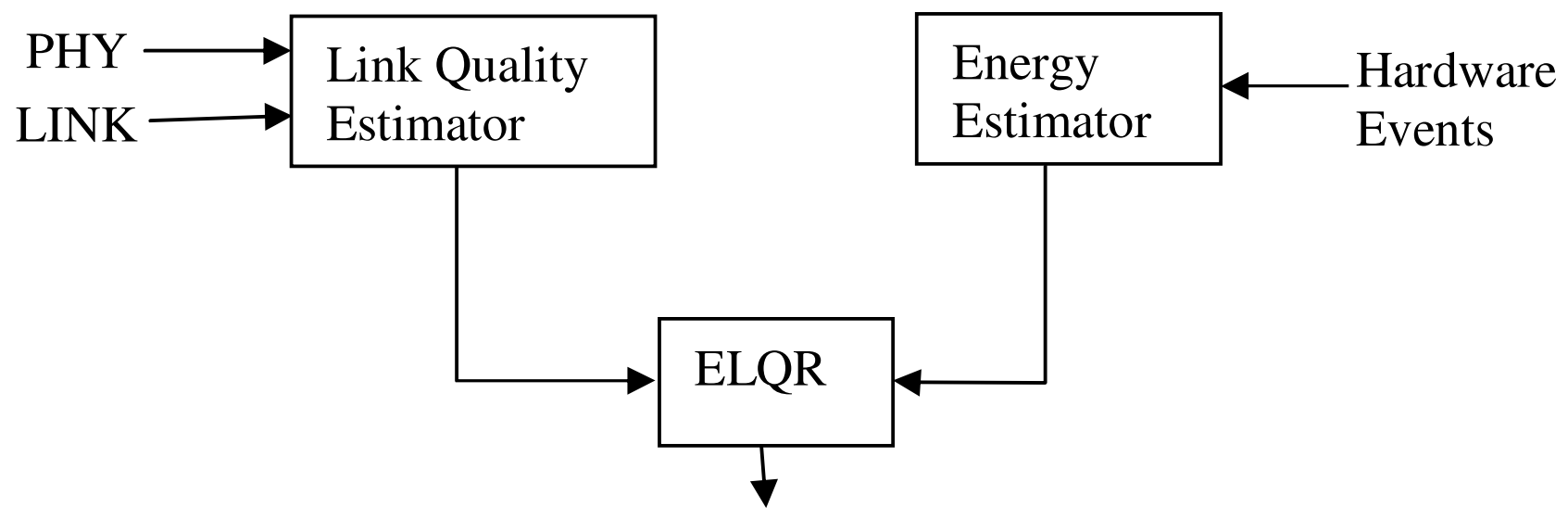

Figure 1. Architecture of EQLR

CTP is best effort tree-based anycast collection tree protocol used to collect the data from various sensor nodes deployed in the field. This protocol is intended for low traffic rates. Some of the nodes in the network advertise themselves as root and the other nodes hearing this beacon will send their data to this root via multihop. The tree is maintained by sending beacons periodically and the beacon message consists of node's parent, hop distance and the cumulative path cost [13].

In four-bit link quality estimation, the nodes will decide their parent based on the routing gradient, which is ETX (Expected Transmissions). ETX measures the expected number of transmissions including the retransmissions required to deliver a packet to its destination. The ETX value for the root is zero and that of other nodes is the sum of ETX value of its parent and the bi-directional value of the link to that parent. The CTP Routing Engine will choose the parent, which has minimum ETX value, i.e. the path with minimum number of transmissions required to transmit a packet to its destination.

Four-bit link estimation also includes the information from the physical, link and network layers to get the good estimation of the link quality [5]. The physical layer can give the information about the channel quality in terms of number of bit errors in the received packets. It sets the white bit when the packets received from that channel are having less bit error, i.e. the channel quality is high during the reception of the packet. The ETX value is estimated by combining broadcast and unicast ETX estimates. For unicast packets, the ETX estimate is the ratio of number of packets transmitted to the number of packets acknowledged. If no packets are acknowledged then the ETX estimate is number of unacknowledged packets from last successful delivery. The ETX estimate for the broadcast packets uses windowed exponentially weighted moving average(EWMA) over the calculated reception probabilities. The broadcast and unicast ETX estimates are combined by EWMA to give ETX value of the link. The network layer decides the best routes based on the information provided by physical and link layer

In our algorithm, the routing table has additional entry of neighbour node's residual energy and the beacon message also carries this information. The algorithm searches for the node with the highest residual energy and the node with minimum ETX from the routing table. If the node has high energy and minimum ETX, then that node will be taken as the parent node. If the nodes are not satisfying any one of the requirements like high energy or low ETX, then the node with maximum energy with ETX value below β threshold will be considered as the next hop or the minimum ETX with energy above α threshold will be considered as the next hop candidate. If none of these conditions are met, the next optimum energy/ETX nodes are considered by searching the table again. The node having highest energy but bad wireless link quality or the node having low energy but good wireless link quality is eliminated from the forwarding list. If the node with high link quality and having less energy is taken as the forwarding candidate, it will create a hole in the network and leads to network partitioning. Similarly, if the node having more energy, but having bad link quality is considered as traffic forwarder, it needs more number of retransmissions and lead to unnecessary energy waste. Hence these nodes are not considered as forwarding candidates. This algorithm is given below.

**1. Initialize()**

```
maxEnergy ← 0
minETX ← 0xFFFF
β ← 50
```

**2. RouteSearch()**

```
for RoutingTable[i]
  If((maxEnergy < RoutingTable[i].energy) & (RoutingTable[i].valid) )
     maxEnergy← RoutingTable[i].energy
     bestEnergyRoute←RoutingTable[i].nodeid
  If(minETX > RoutingTable[i].ETX)  & (RoutingTable[i].valid) )
     minETX ← RoutingTable[i].ETX
     bestETXRoute ← RoutingTable[i].nodeid
```

**3. ParentSelection()**

```
//Choose the link with high energy and good quality link
If (bestETXRoute == bestEnergyRoute)
      Parent = bestEnergyRoute.nodeid

//Choose the best ETX path
```

```
    elseif (bestETXRoute.Energy > α)
        Parent ← bestETXRoute.nodeid
//Choose the best Energy path
    elseif (bestEnergyRoute.ETX < (minETX +β))
        Parent ß bestEnergyRoute.nodeid
        β = β + β * Round/ 100
// search for the next alternative parent
    elseif
        bestEnergyRoute.valid = 0;
        bestETXRoute.valid = 0;
        Repeat step 1
```

Figure 2 Algorithm for ELQR

## 4. IMPLEMENTATION OF ELQR

The challenges involved in the development of energy based routing protocol for the sensor node is the exact measurement of energy. The hardware platform such as IRIS from Crossbow Technologies, USA does not provide any hardware based energy measurement and hence the energy estimation should be done using software method [14]. The energy used by the sensor mote is calculated at run time by tracking the time spent in different operating modes by the different hardware components such as microcontroller, radio, LED, sensors and memory. Then the energy consumption of these components is calculated by using the energy budget of the IRIS mote provided in the user manual [15]. The residual energy of the mote is estimated using linear energy model at run time and this is updated in the neighbour table of the mote [16].

The routing engine CtpRoutingEngine.nc is modified to implement the proposed energy and link quality based routing protocol ELQR. The new interface *Energy* provides the command, which returns the residual energy of the mote and this is called periodically by the routing engine. The energy value is sent with the beacon message and the nodes, which receives beacon, will update their neighbour table with the updated energy value. The routing decision is modified according to the given algorithm in the U*pdateRoute* task. The beacon interval is more and hence the energy of the parent in the neighbour table won't be the current value. To have the more updated value, the residual energy of the mote is also sent in the data packets, so that the neighbouring nodes receiving or snooping the packets will update their neighbour table. Therefore the neighbours will have the most updated energy value of its parent.

The energy threshold α is set to 14,400 joules, which corresponds to 2V. From the IRIS data sheet and the test conducted at our deployment field, the mote operating range is from 1.8V to 3V. The mote will not communicate if the battery voltage goes below 1.8V. Therefore the threshold must be set just above 1.8V. The capacity of the battery is calculated using the formula $V * A_b$, where V is the supply voltage in volts and $A_b$ is the current drawn per hour in Amp – hour. The IRIS motes operate on 2 AA batteries and the current capacity is 2200 mA-hour. Hence the threshold battery capacity at 2V is found as 2 * 2.2 * 3600 = 14,400 Joules.

This algorithm always chooses the best ETX path until the residual energy of the node is above the energy threshold. As the nodes are continuously used for forwarding the packets, the energy of the node goes beyond the threshold value, and is eliminated from the routing table. The energy threshold is set such that they can be alive in the network without involving in the routing task. The new routes will be found by taking the next best energy routes with ETX value below the threshold β. The ETX value is a cumulative path cost, which also depends on the number of hops from the sink. If the threshold β is changed, the route selected may be having more number of hops than the one with best channel characteristics. The threshold β is changed from 50 to 500 as the data collection round increases. The β value is increased for every 100 data collection round, enables the routing protocol to select the longer route with maximum energy. The threshold values are chosen from our test bed consists of 9 motes and the maximum value measured for ETX is 500.

The performance of the algorithm is tested with TOSSIM simulator and the same is verified using the IRIS test bed.

# 5. Performance Analysis

## 5.1 Simulation using TOSSIM

The performance of the proposed protocol is tested extensively using TOSSIM under Linux, which is a simulator for TinyOS developed by University of California at Berekely, USA, which can run the actual TinyOS code without any real motes [6]. The hardware components are simulated using software at packet level and the code developed for this simulator can be directly fused into the motes with minimum changes.

The simulation parameter used for creating the network topology is given below. The nodes are placed in uniform topology where the area is divided into grid of equal size and the node is placed randomly inside the grid. The channel parameters used in this simulation depicts the indoor environment and the nodes are having highly asymmetric links.

Table 1. Simulation Parameters

| **Parameter** | **Value** |
|---|---|
| No. of Nodes | 9 |
| Topology | Uniform, Random |
| Size | $50 \times 50$ |
| Radio Model | Indoor |

The network consists of 9 nodes, which are placed randomly in the sensor field, and generates traffic for every 1 second. The sink collects the packet and forwards the data to the PC. The simulation time varies from 1000 simulation seconds to 10000 simulation seconds. The number of packets forwarded by each node, total packets sent by each node and the total packets received by the sink are calculated. The figure 3 shows the network topology of 9 nodes and the node 0 is the sink, which is connected to the PC.

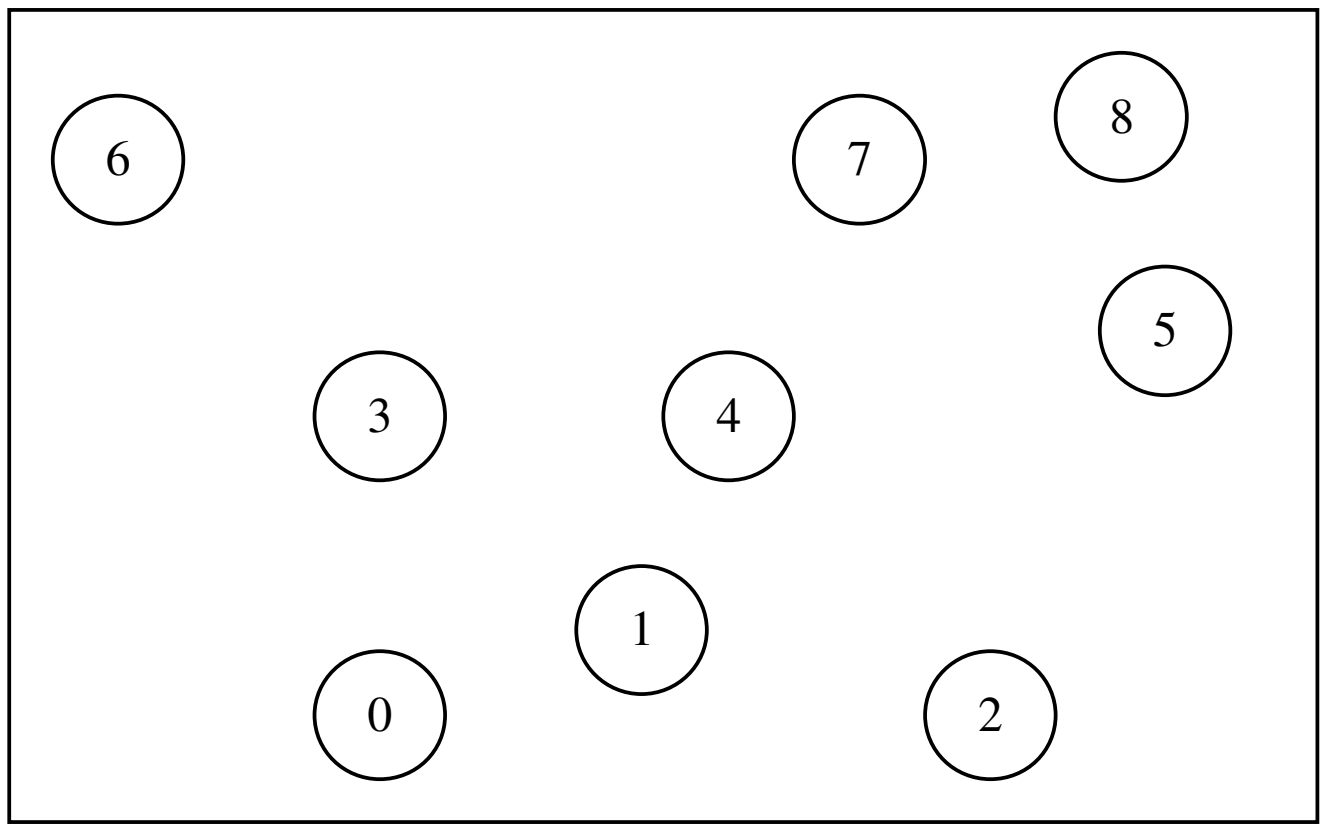

Figure 3. Network Topology

In order to analyse the performance of the routing protocol, the parameters *Packet Reception Ratio (PRR)* and the *load balance* of the networks has been chosen.

- *Packet Reception ratio* is the ratio between the total numbers of packets received by the sink to the total number of packets transmitted by all the nodes for a specific duration.
- *Load balance or Load distribution* is the number of packets forwarded by each node during a fixed duration of time.

The figures 4 to 8 show the load distribution for various simulation times.

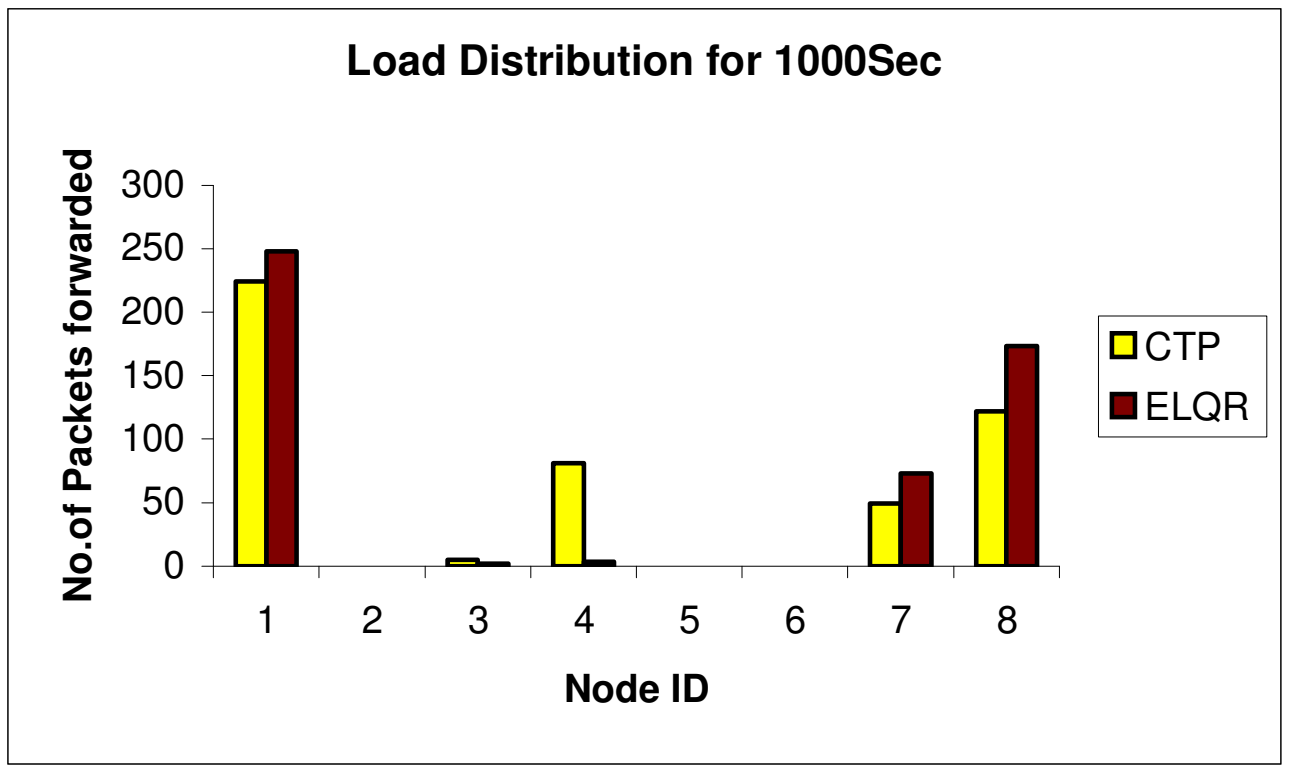

Figure 4. Load Distributions for 1000 Seconds

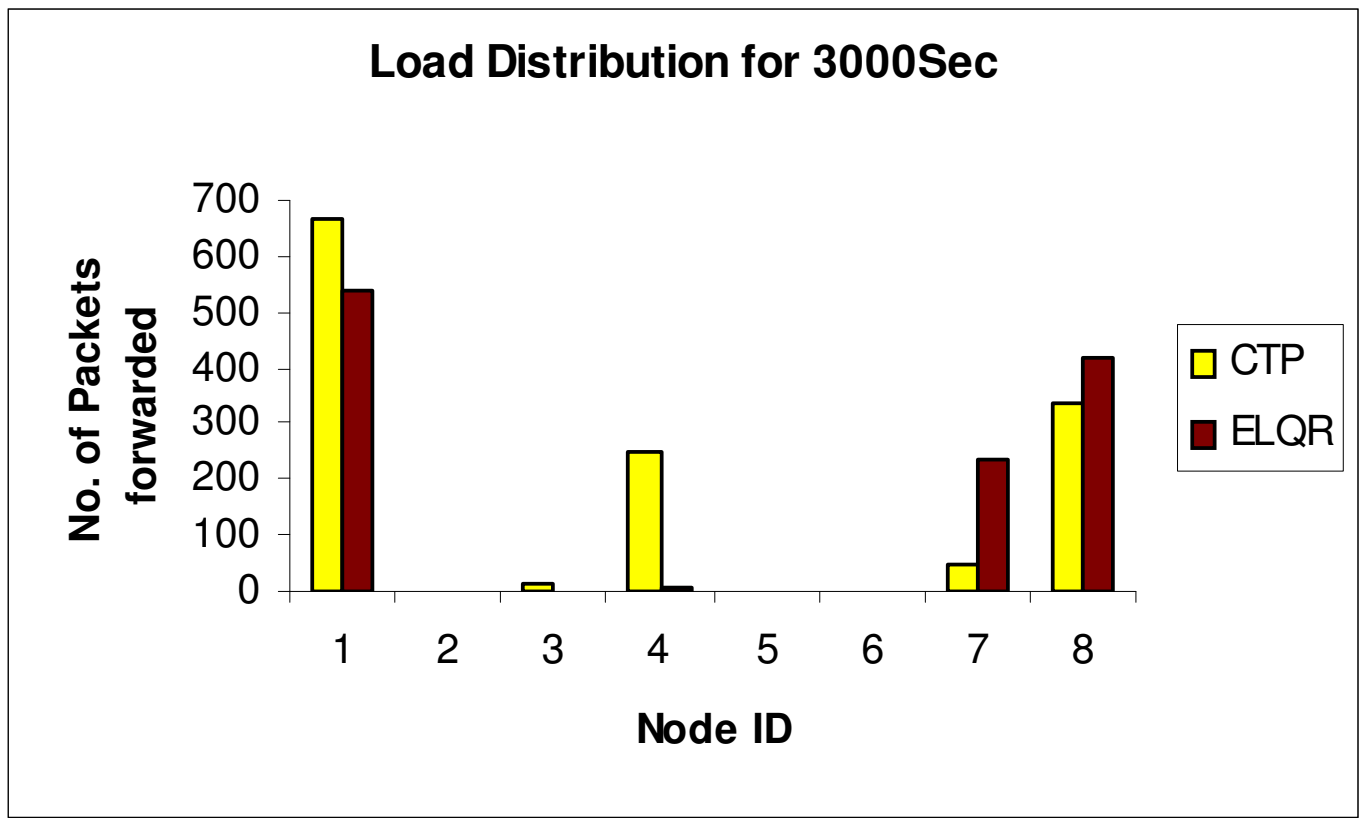

Figure 5. Load Distributions for 3000 Seconds

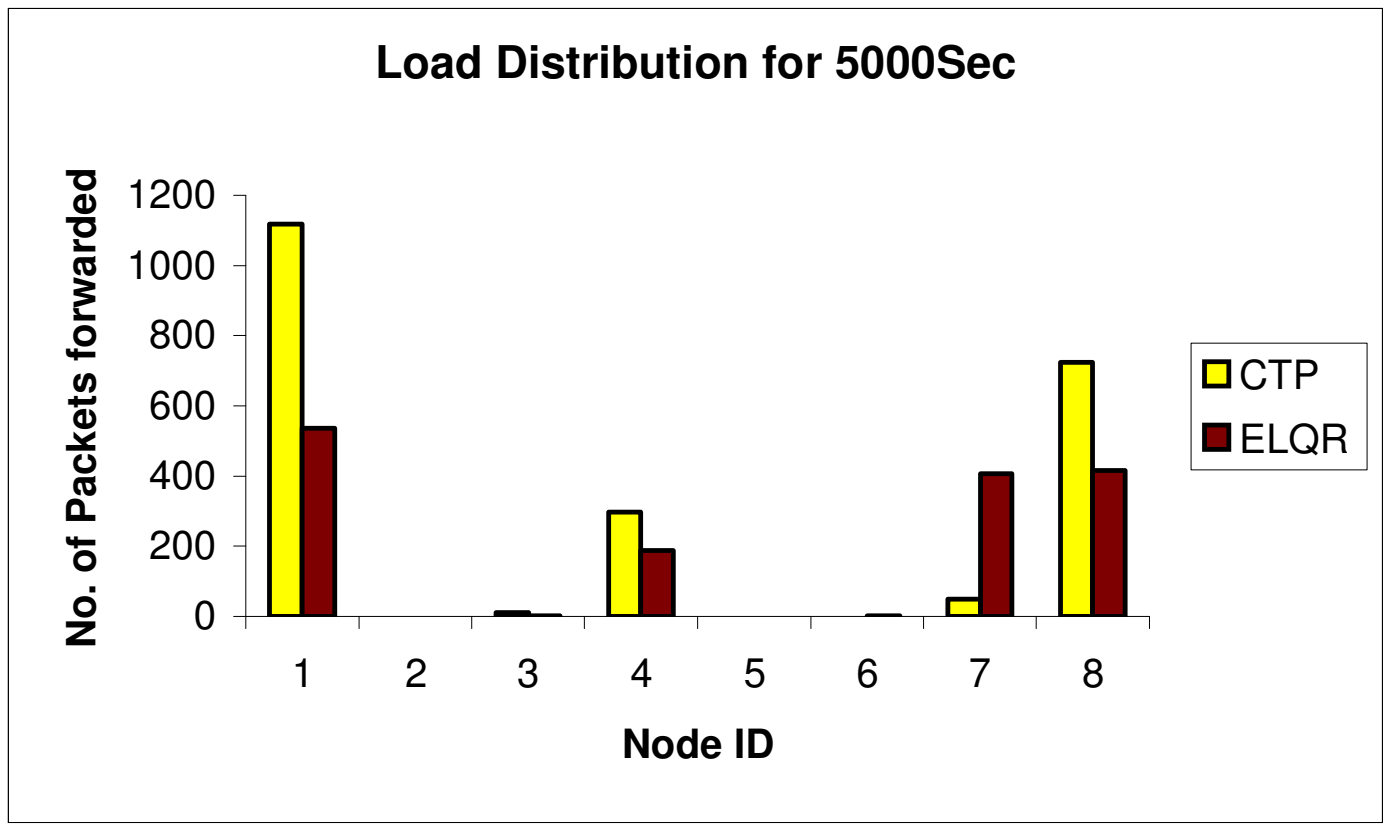

Figure 6. Load Distributions for 5000 Seconds

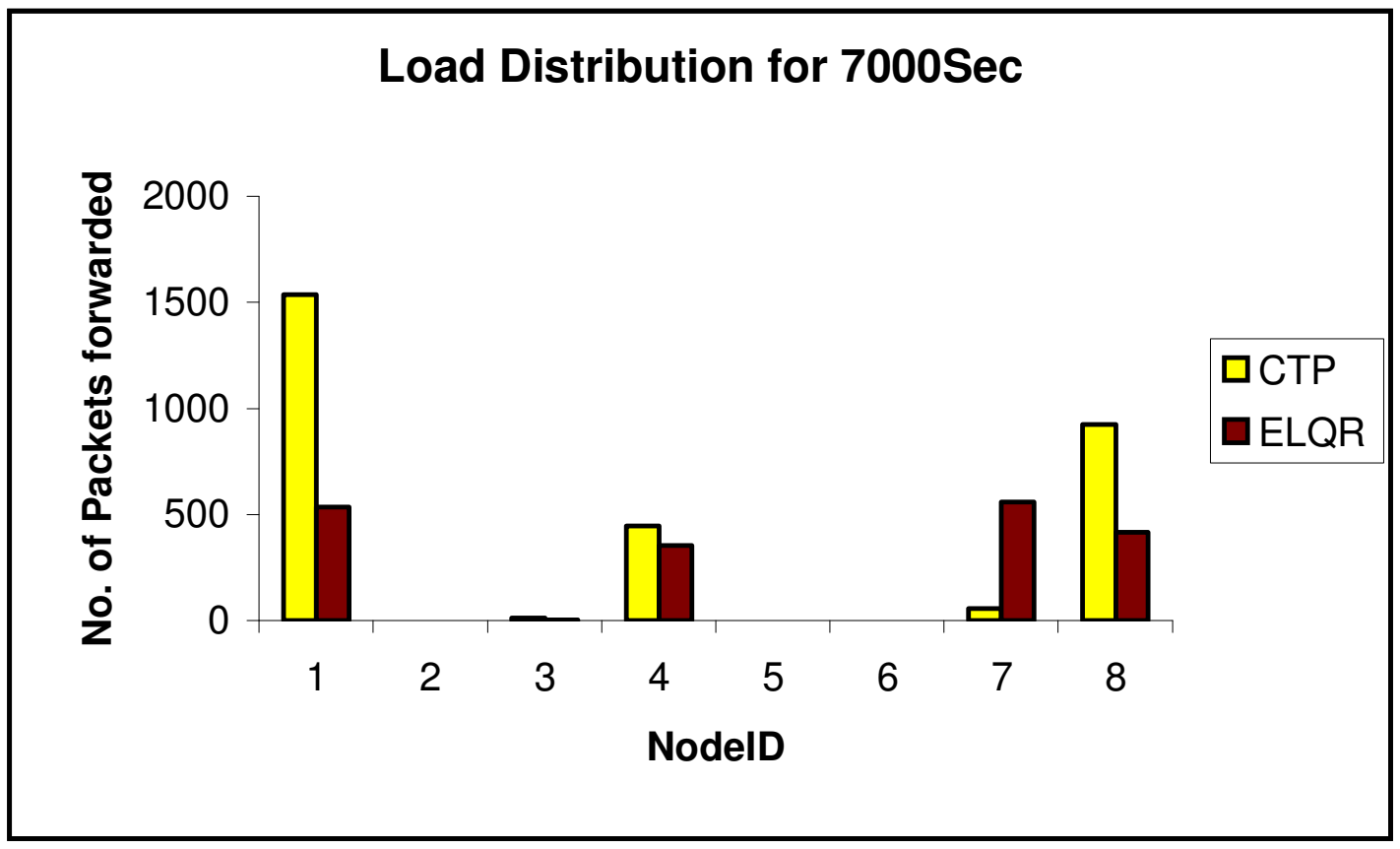

Figure 7. Load Distributions for 7000 Seconds

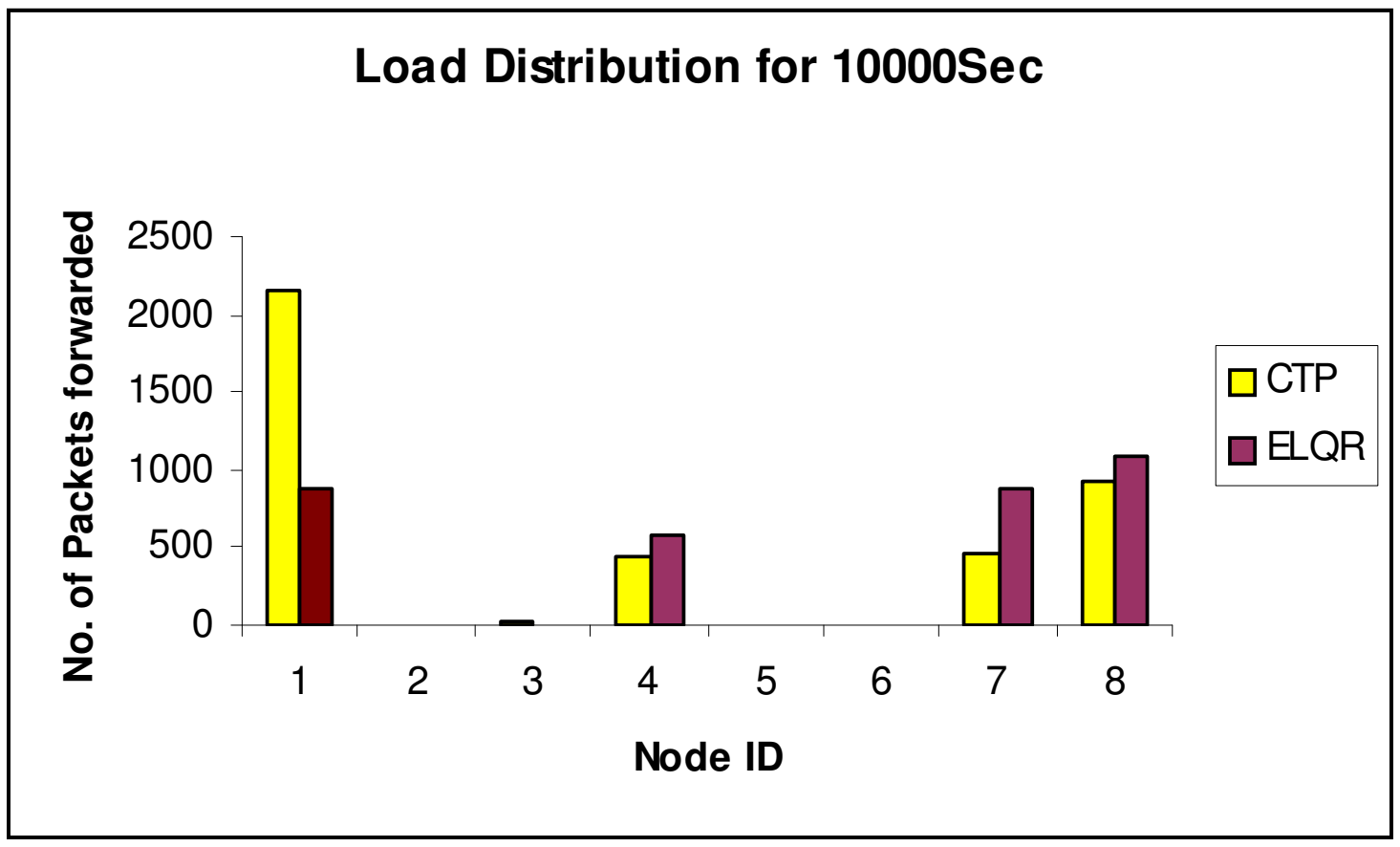

Figure 8. Load Distributions for 10000 Seconds

The graphs show the total number of packets forwarded by each node. The packet of its own is not included in the calculation, since the packets are transmitted at periodic interval of time. The nodes 1, 4, 7 and 8 are intermediate nodes and they forward the packets from its neighbours. These nodes involve more communication activities and hence they may be drained out soon. The node 1 is near to the sink and forwards more packet in the default routing protocol CTP. From the figure 4 and 5, it is understood that the load distribution among the nodes in both the protocols are similar. As time increases, CTP uses the same nodes as forwarding candidates and hence the drain out happens soon. But in ELQR, the traffic load of node 1is shared with that of node 4 and node 8's load is reduced by diverting the traffic to node 7. From the figures 6,7 and 8, it is understood that in ELQR, the nodes will take different route to reach the sink and hence the lifetime of the network is increased. Also, the forwarding loads are almost equal in the case of ELQR where as in CTP, few nodes forward more packets.

The figure 9 shows the PRR value for these two protocols for the simulation times from 1000 to 10000seconds. The PRR value of the proposed scheme is less compared to the CTP protocol. The PRR value of the EQLR protocol is almost same as that of CTP initially, but decreases as time advances. This may be due to the searching of new parent if the current parent is not satisfying the routing criteria.

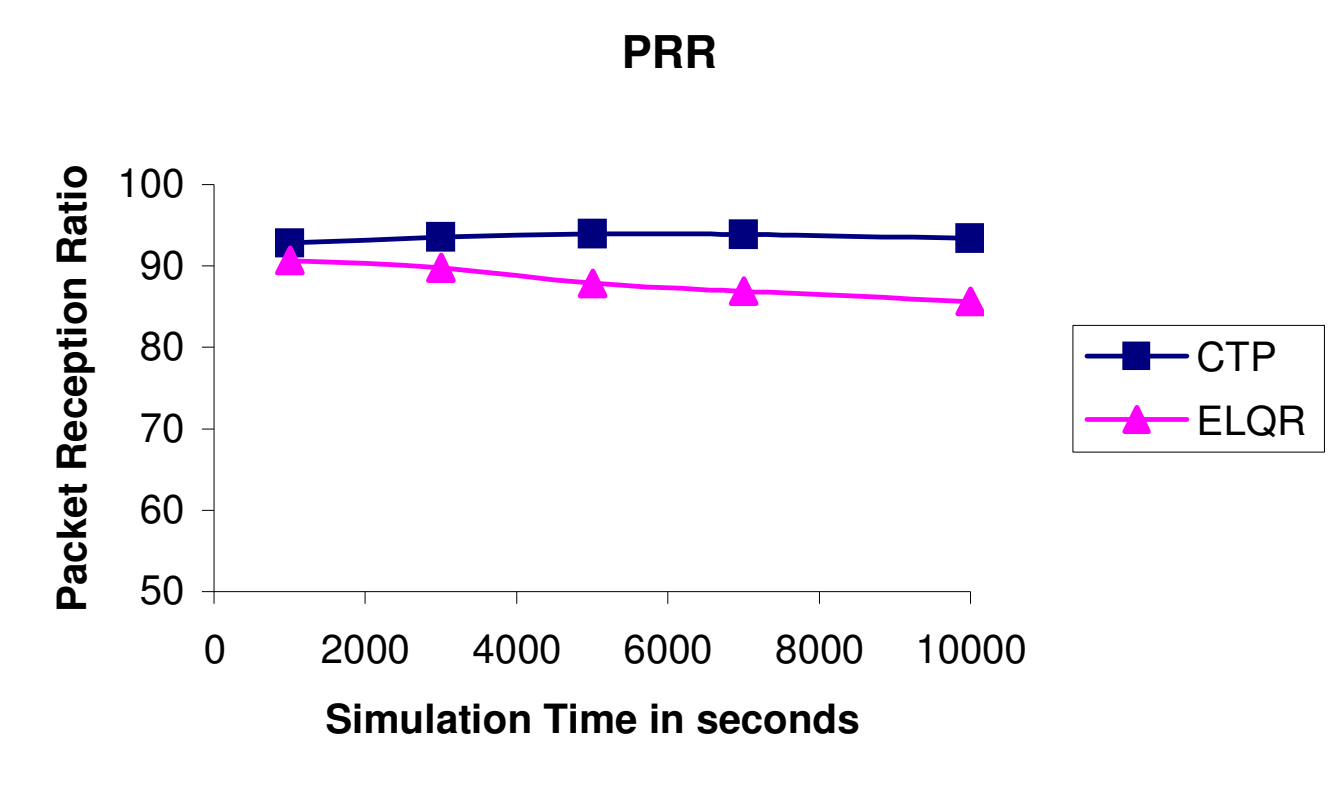

Figure 9. Packet Reception Ratio

The performance of the protocol is tested for 100 nodes, which are placed 500 × 500 area with same simulation parameters as given above. The number of nodes alive over the simulation time is measured and is shown in the figure 10. The time at which the first node dies, is also calculated for CTP and ELQR and are 2487 and 3950 seconds respectively.

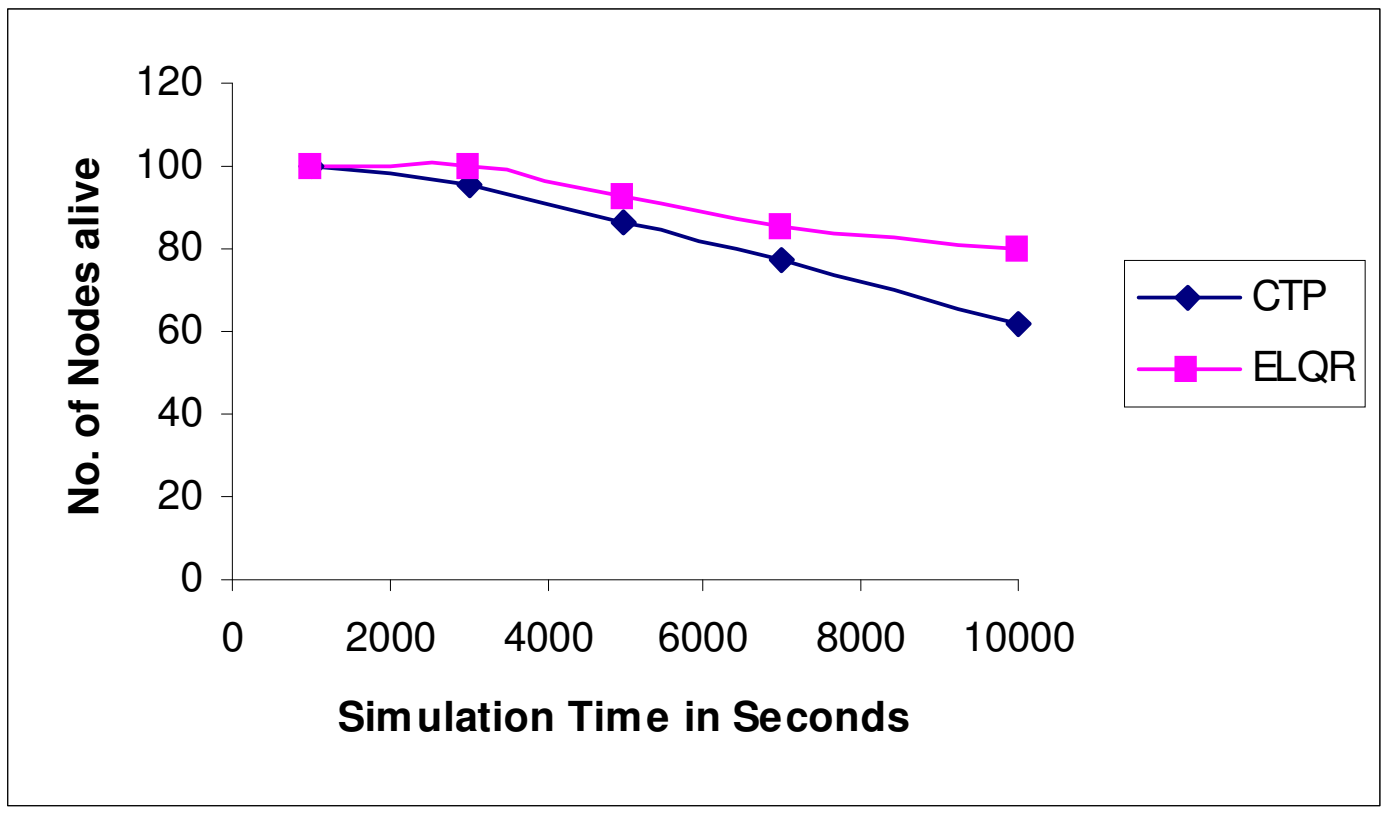

Figure 10. Number of nodes alive

## 5.2 Performance of ELQR in IRIS Test bed

The test was conducted in our lab for an hour and the nodes generates the data every one second. The application program *MultihopOscilloscope* reads the temperature sensor of the MTS300 sensor board and the temperature value is sent to the sink for every seconds. The motes are initially using the default routing protocol CTP and the number of packets sent, forwarded by each node is measured. The total number of packets received by the sink is also measured. The table 2 gives the measured readings taken from the test bed.

Table 2. CTP

| Node ID | Sent | Forwarded | Received | Parent |
|---|---|---|---|---|
| 0 | - | - | 9554 | 0 |
| 1 | 788 | 5481 | | 0 |
| 2 | 1239 | 30 | | 0 |
| 3 | 1256 | 3250 | | 1 |
| 4 | 1212 | | | 3 |
| 5 | 789 | 1134 | | 1 |
| 6 | 1265 | | | 5 |
| 7 | 1224 | 1097 | | 3 |
| 8 | 1127 | | | 7 |

From the table, the node 1 and 3 involves in more communication. The node 5 may change the route between the node 2 and node 1 due to change in the channel characteristics. But mostly the node 5 chooses the node 1 as its parent and it sends more packets to node 1. The tree formed using CTP algorithm is shown in the figure 11. The topology changes whenever there is an disturbance in the channel due to human intervention.

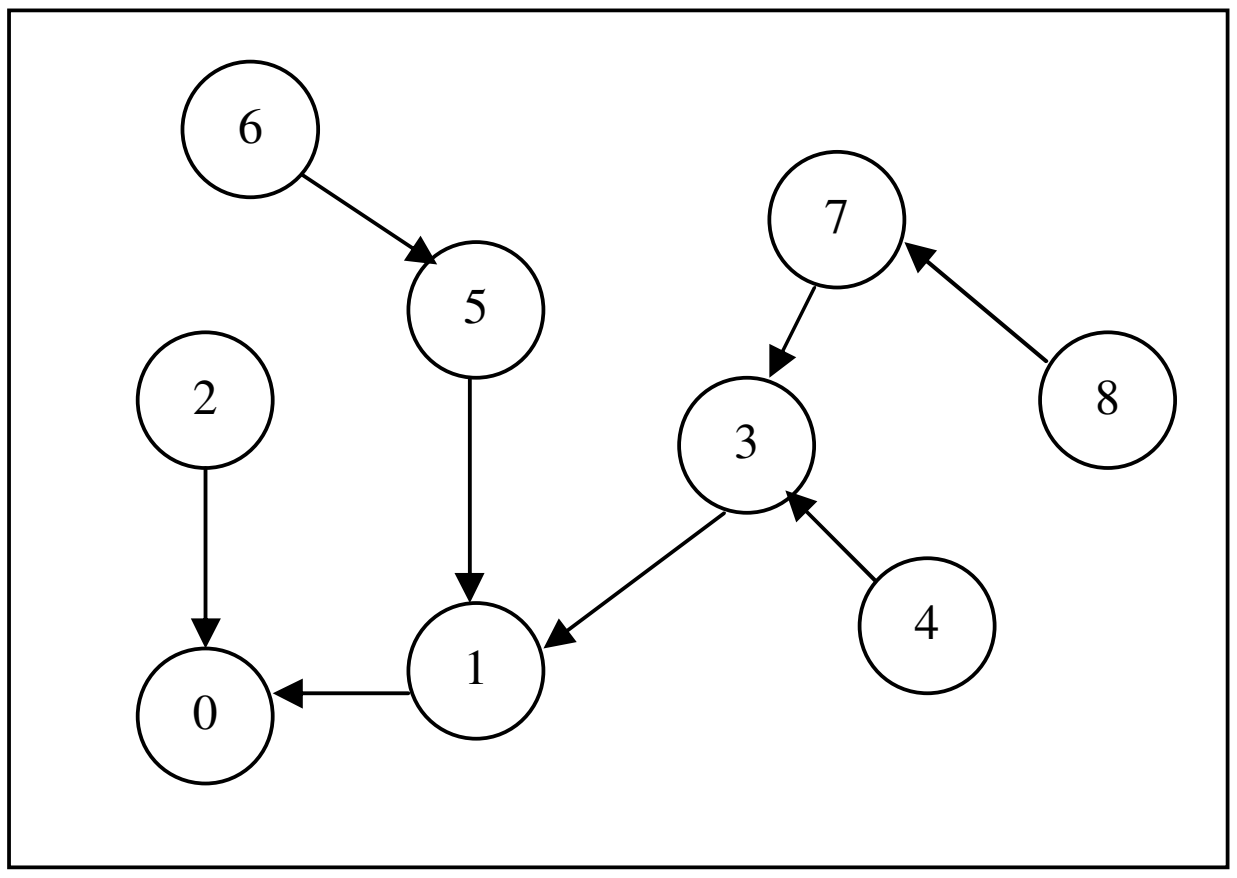

Figure 11. Tree using CTP

The proposed algorithm ELQR is loaded on the motes and they are placed at same location as in CTP. The same testing environment is maintained and the similar parameters as CTP are calculated. This is given in the table 3.

Table 3. ELQR

| Node ID | Sent | Forwarded | Received | Parent |
|---|---|---|---|---|
| 0 | - | - | 9671 | 0 |
| 1 | 1135 | 3509 | | 0 |
| 2 | 1250 | 1585 | | 0 |
| 3 | 1138 | 3056 | | 5 |
| 4 | 1267 | | | 3 |
| 5 | 1241 | 2364 | | 2 |
| 6 | 1198 | | | 5 |
| 7 | 956 | 1097 | | 3 |
| 8 | 1173 | | | 7 |

From the table 3, the node 1 traffic is reduced by 36% in ECTP as compared to CTP. The nodes 3 and 5 changed their parent from node 1 to node 5 and 2 respectively. This decision was made based on the residual energy of the node and hence the node 1 lifetime is extended. The initial topology of the ELQR protocol is same that of CTP which is shown in the figure 11. But after some time, the topology is changed based on the energy. The topology of the ECTP after 1 hour is given below.

Further the test was extended to find how long the sink could receive the data from the network. All the motes are energized with the new battery. It is found that the network using CTP was alive for 18hours 20 minutes and the network using ELQR can work upto 21 hours and 40 minutes.

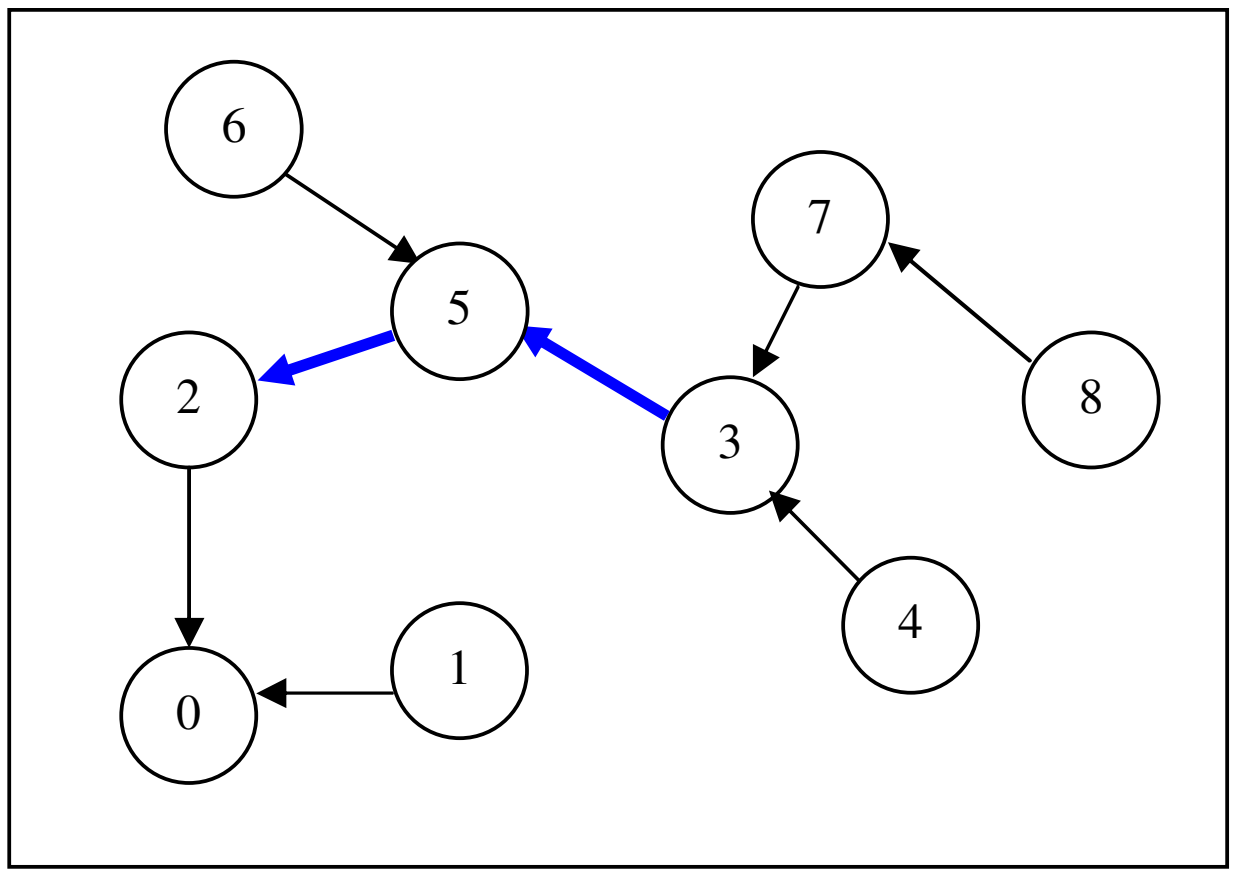

Figure 12. Tree using ELQR

## 6. CONCLUSION

Energy is an important resource in WSN and to enhance the lifetime of the network, the traffic load should be distributed among the forwarding nodes such a way that they could be alive for more time. In this paper, we proposed a protocol, which takes the routing decision based on the link quality and the residual energy of the nodes. Each node will make one of its neighbours as its parent if the neighbour node has sufficient energy to forward the packet and the quality of the link to that node is good. A threshold is said for both parameters and the energy threshold is set well above 2V such that the lifetime of the node will be increased by reducing its load. The link quality increases with hop count and the node with less value will be chosen as the forwarding candidate. In this work, the threshold for the link quality is changed through out the lifetime of the network to make the network alive by selecting the longer routes with more energy. Hence the routes with more energy and adequate wireless link quality will be chosen for the forwarding operation with out exhausting the nodes in the optimal path. This algorithm is simulated on TOSSIM and it is tested on the test bed comprising of IRIS motes. This algorithm performs well compared to CTP by extending the network lifetime. The packet reception ratio is less compared to CTP due to the fact that the convergence time of the algorithm is more when there is a change in the route. The effect on change in energy threshold will have the impact on the performance of the network and it could be explored.

**Authors**

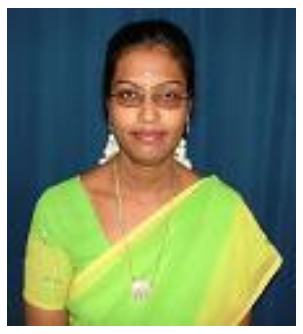

**A.Sivagami** received her B. E. degree in Electronics and Communication Engineering from University of Madras in the year 1990 and obtained her Master's degree in Communication Systems from Bharathiyar University in the year 2000. Currently she is doing her research in the field of Wireless Sensor Networks in Anna University, Chennai, India. Her fields of interests are Wireless Networks, Wireless Sensor Networks, Digital Communication, etc She has published 3 papers in International conferences and 2 in International Journals.

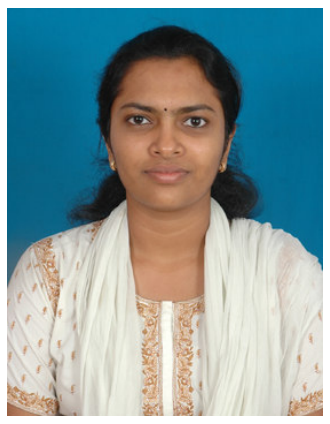

**K.Pavai** received her B.E. degree in Electronics and Instrumentation Engineering in the year 2004 from university of Madras and M.E. in Applied Electronics in the year 2006 from Anna University, Chennai. Currently she is doing her research in the field of Wireless Sensor Networks in Anna University, Chennai. She has 4 publications in the National conferences, 3 in International conferences and two in International Journals. Her current field of interests are Wireless Sensor Networks, VLSI, Mobile Communication, Instrumentation, etc.

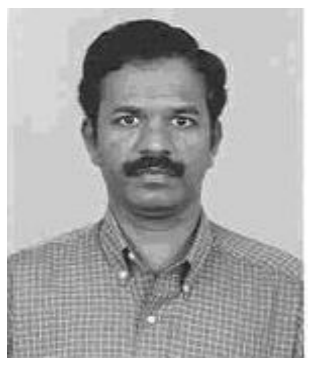

**Dr.D.Sridharan** received his B.Tech. Degree in Electronics Engineering and M.E. degree in Electronics Engineering from Madras Institute of Technology, Anna University in the years 1991 and 1993 respectively. He got his Ph.D degree in the Faculty of Information and Communication Engineering, Anna University in 2005. He is currently working as Assistant Professor in the Department of Electronics and Communication Engineering, CEG Campus, Anna University, Chennai, India. His present research interests include Internet Technology, Network Security, Distributed Computing and VLSI for wireless Communications. He has published more than 25 papers in National/International Conferences and Journals.

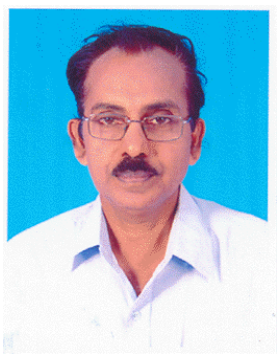

**Shri S. A. V. Satya Murty** received his B.Tech degree from Jawaharlal Nehru Technological University in 1977. He has completed one-year orientation course in Nuclear Science and Engineering at Bhaba Atomic Research Centre, Bombay. He is the Head of Computer Division, IGCAR, Kalpakkam. He played key role in the establishment of Mainframe Computer Systems, Internet and E-mail facilities, High Performance Computing facility, Intra DAE VSAT Networks, Grid Computing Facility, etc. He is currently working on PFBR Simulator, Wireless Sensor Networks, Computational Intelligence, Advanced Visualization Centre, and Knowledge Management etc. His research interests are Network Security, Sensor Networks, data base management and knowledge management, etc. He has published more than 50 technical publications in refereed journals and conferences.